\documentclass[12pt]{article}

\usepackage[margin=1in, top=1in, bottom=1in]{geometry}

\usepackage{amsmath,amssymb,amsthm,epsfig,rotfloat,psfrag,natbib,url,graphicx,lineno}
\usepackage[color=white]{todonotes}

\usepackage{lmodern}

  \newcommand{\by}{\mathbf{y}}
  \newcommand{\bt}{\boldsymbol{\theta}}
  \newcommand{\bvt}{\boldsymbol{\vartheta}}
  \newcommand{\bb}{\boldsymbol{\beta}}
  \newcommand{\bn}{\boldsymbol{\eta}}
  
  \newcommand{\bp}{\boldsymbol{\psi}}
  
  \newcommand{\bg}{\boldsymbol{\gamma}}

  \newcommand{\bm}{\boldsymbol{\mu}}
  \newcommand{\bSig}{\boldsymbol{\Sigma}}
  
  \newcommand{\ba}{\boldsymbol{\alpha}}
  \newcommand{\bxi}{\boldsymbol{\xi}}
  \newcommand{\bd}{\boldsymbol{\delta}}
  \newcommand{\bs}{\boldsymbol{\sigma}}
  \newcommand{\bS}{\mathbf{S}}
  \newcommand{\bL}{\mathbf{L}}
  \newcommand{\bK}{\mathbf{K}}
  \newcommand{\bX}{\mathbf{X}}

  \newcommand{\bV}{\mathbf{V}}
  \newcommand{\bZ}{\mathbf{Z}}

  \newcommand{\bu}{\mathbf{u}}
  \newcommand{\tN}{\text{N}}
  \newcommand{\bY}{\mathbf{Y}}
  \newcommand{\bI}{\mathbf{I}}
  \newcommand{\bz}{\mathbf{0}}

\begin{document}


\vspace*{\fill}

\begin{center}
\setlength{\parindent}{0pt}
\renewcommand{\baselinestretch}{1.8}\normalsize

{\Large Greater Than the Sum of its Parts: \\Computationally Flexible Bayesian Hierarchical Modeling}

\renewcommand{\baselinestretch}{1.15}\normalsize 
\bigskip\bigskip

Devin S. Johnson\footnote{Corresponding author: {\tt devin.johnson@noaa.gov}}\\ 
Pacific Islands Fisheries Science Center,\\
National Marine Fisheries Service, NOAA \medskip

and\medskip

Brian M. Brost\\
Alaska Fisheries Science Center,\\
National Marine Fisheries Service, NOAA \medskip

and\medskip

Mevin B. Hooten \\
Department of Statistics and Data Sciences \\
The University of Texas at Austin

\bigskip\bigskip

\today

\end{center}

\vspace*{\fill}

\clearpage

\renewcommand{\baselinestretch}{1.65}\normalsize
\raggedright
\setlength{\parindent}{2em}



\noindent {\bf Abstract} 
We propose a multistage method for making inference at all levels of a Bayesian hierarchical model (BHM) using natural data partitions to increase efficiency by allowing computations to take place in parallel form using software that is most appropriate for each data partition. The full hierarchical model is then approximated by the product of independent normal distributions for the data component of the model. In the second stage, the Bayesian maximum {\it a posteriori} (MAP) estimator is found by maximizing the approximated posterior density with respect to the parameters. If the parameters of the model can be represented as normally distributed random effects then the second stage optimization is equivalent to fitting a multivariate normal linear mixed model. We consider a third stage that updates the estimates of distinct parameters for each data partition based on the results of the second stage. The method is demonstrated with two ecological data sets and models, a random effects GLM and an Integrated Population Model (IPM). The multistage results were compared to estimates from models fit in single stages to the entire data set. In both cases multistage results were very similar to a full MCMC analysis.

\bigskip





\section{Introduction}

Bayesian hierarchical models (BHMs) have become ubiquitous in many fields of scientific enquiry because of their ability to flexibly model complex natural systems yet remain relatively easy to build and modify for the questions at hand \citep{hobbs2015bayesian}. \citet{berliner1996hierarchical} clarified conceptualization of a BHM based on three submodels: (1) {\it data}, (2) {\it process}, and (3) {\it parameter} models. The data model accounts for measurement error or other differences between observable data and the true natural process for which the researcher would like to make inference. The process model describes the state of nature using a random process to account for its unknown intricacies. Finally, the parameter model describes our uncertainty about the parameters governing this random process.  The full model is built by conditioning each submodel on the one above it. Traditionally, inference for BHMs has been conducted using some form of Markov Chain Monte Carlo (MCMC) because the model structure is ideally suited to it \citep{gelfand1990sampling,gelfand2015hierarchical}. If not for development of MCMC methods, widespread use of BHMs may never have come to pass \citep{green2015bayesian}. With a large amount of data or complex submodels, however, MCMC can be computationally challenging or infeasible to implement \citep{hooten2018prior, wikle2003hierarchical}. Herein, we propose a multistage method for making inference at all levels of a BHM using various software formats in each stage. By breaking the estimation procedure into multiple stages based on a partition of the full data it becomes computationally efficient. By allowing use of the most suitable software platform in the initial stage it can be made more efficient yet, not only in computing time, but also in development time saved by using software designed to facilitate each specific initial stage analysis. 

To alleviate computational challenges in using MCMC to fit BHMs in ``big data'' situations there have been several avenues of approach  developing in the literature. One of these is development of so called ``2-stage'' methods that seek to break the computation into pieces based on partitions of the data \citep{goudie2019joining, hooten2016hierarchical,hooten2018prior,lunn2013fully,mesquita2020embarrassingly,scott2016bayes}. The first stage involves fitting separate data models to the partitions using MCMC, then in the second stage, the separate MCMC samples are combined to produce an MCMC sample as if the partitions had been simultaneously analyzed with the full BHM. A similar approach is the divide-and-conquer approach \citep{srivastava2018scalable} averages posterior distributions from data partitions to approximate an analysis of the full data. 

A alternative approach to default MCMC for computationally challenging BHMs is to use optimization methods for exploring the posterior parameter space rather than stochastic sampling \citep{green2015bayesian}. Optimization of the posterior density aims to find the  {\it maximum a posteriori} (MAP) estimate rather than the posterior mean for a point estimate. The main benefit is that optimizing the posterior with respect to the parameters often requires fewer evaluations of the posterior density. That comes at a cost, however, because asymptotic results are usually required to describe parameter uncertainty \citep{van2000asymptotic}. 

In an effort to provide a computationally efficient and flexible method for fitting a general class of BHMs, {\it conditionally independent hierarchical models} (CIHMs; \citealt{kass1989approximate}), we consider combining benefits of meta-analysis, 2-stage MCMC, and posterior optimization into a multistage method that can be used to quickly fit BHMs in big data situations or when the data-level models are complex themselves. Although meta-analysis and 2-stage MCMC are both accomplished in multiple stages, 2-stage MCMC was initially designed to analyze a BHM, whereas meta-analysis seeks to summarize results from various analyses by forming them into a BHM, namely, a multivariate normal linear mixed model \citep{gasparrini2012multivariate}. We can use the meta-analysis approach to approximate the second stage combination of initial stage inference in a 2-stage MCMC. This allows us to use multiple approaches  to approximate the posterior inference in the initial stage, including optimization methods, deterministic sampling, or MCMC if it is convenient. 

In what follows, we provide notation and a working definition of the CIHM as we will use it. We show how the full posterior can be approximated using multivariate normal approximations to first stage posterior densities. Then we demonstrate how the first-stage results can be combined using the linear mixed model approach of standard meta-analysis \citep{gasparrini2012multivariate,higgins2009re} when the first-stage parameters are random effects. An optional third stage is introduced to retroactively revisit the first stage analysis and can improve estimates after the second-stage results are obtained. The method is then demonstrated for two different types of environmental data, a simple generalized linear mixed model (GLMM) and an integrated population model (IPM) for demographic parameter inference. 

\section{Conditionally independent hierarchical model}

The description of BHMs by \cite{berliner1996hierarchical} is very general and we only consider a specific, but still very broad, class of hierarchical models (CIHMs; \citealt{gelfand2015hierarchical, kass1989approximate}). The general form of a CIHM is
\begin{equation}\label{eq:full.model}
\begin{aligned}
\text{Data-level model: } & \by = (\by_1',\dots,\by_n')' \sim \prod_{i=1}^n[\by_i|\bt_i, \bg_i, \bn], \\
\text{Unit-level effects: } & \bt = (\bt_1',\dots,\bt_n')' \sim [\bt|\bp],\\
 & \bg = (\bg_1',\dots,\bg_n')' \sim \prod_{i=1}^n[\bg_i], \\
\text{Population effects: } & (\bp',\bn')' \sim [\bp,\bn],
\end{aligned}
\end{equation}
where ``$[A|B]$'' is used to represent the conditional probability density or distribution function of $A$ given $B$ \citep{gelfand1990sampling}, $\by_i$ is the observed data set (or data partition) for the $i$th unit ($i=1,\dots,n$), $\bt_i$ are random unit-level parameters, $\bp$ are population-level parameters for $\bt_i$, $\bg_i$ are distinct (relative to each $\by_i$) unit-level parameters for modeling the $i$th data set, and $\bn$ are global (fixed) population-level parameters. Typically, $\bp$ and $\bn$ are the most scientifically interesting, however, there can be interest in each of the $\bt_i$ and $\bg_i$ depending on the situation. Note, that unit-level effects are allowed to be correlated which is slightly more general than \citet{kass1989approximate}. 

To place the development of the multi-stage methodology in context, we will demonstrate its use on the analysis of two ecological data sets. The first one uses a generalized linear mixed model (GLMM) to assess the effect of wolf ({\it Canis lupis}) presence on the  occurrence of moose ({\it Alces alces}) browse in willow stands. Although these data are not large in the modern sense, we demonstrate a 2-stage analysis of them because the model can easily be fit in its full hierarchical form for comparison. Another common group of models in the ecological literature are integrated population models (IPMs). Initially proposed by \cite{besbeas2002integrating}, IPMs are routinely used to collect information in multiple sparse data sets for improved inference on shared parameters. To demonstrate the 2-stage meta-analysis approach for IPM model fitting, we analyze the classic northern lapwing ({\it Vanellus vanellus}) data from \cite{besbeas2002integrating}.

\subsection{Moose browse in the presence wolves (GLMM)} \label{sec:moose.data}

This example involves $n =$ 2,914 binary responses, $y_{ij}$, which indicate presence of moose browsing on tree $j=1,\dots,n_i$ of plantation $i=1,\dots,24$. The original study assessed whether or not tree browsing is reduced in plantations exposed to high wolf activity \citep{van2018does}. In addition, the study accounted for inherent moose preferences for tree height. To perform a similar analysis, we use a random effects model to treat each plantation as a unit to learn about the fixed effect of high wolf presence on moose browsing while accounting for natural variation in moose response across the 24 plantations. The CIHM is given by
\[
\begin{gathered}
y_{ij} \sim \text{Bernoulli}(p_{ij}), \\
\text{logit}(p_{ij}) = \delta_0 + \delta_{\text{height}}h_{ij} + \delta_{\text{wolf}}w_i + u_{0i} + u_{\text{height},i} h_{ij},\\
\mathbf{u}_i = (u_{0i}, u_{\text{height},i}) \sim \text{N}(\mathbf{0},\text{diag}(\sigma_1^2, \sigma_2^2)),
\end{gathered}
\]
where $h_{ij}$ equals the height of the $j$th tree in the $i$th plantation and $w_i$ is a binary indicator of high wolf presence in the $i$th plantation.

To use the 2-stage approach we must hierarchically center the model because $\delta_0$ and $\delta_{\text{wolf}}w_j$ are not identifiable within plantation units. Using hierarchical centering \citep{Gelfand:1996zq}, the model can be rewritten as,
\[
\begin{gathered}
\text{logit}(p_{ij}) = \theta_{0i} + \theta_{\text{height},i}h_{ij},\\
\bt_i = (\theta_{0i}, \theta_{\text{height},i}) \sim \text{N}(\mathbf{L}_i\bd, \bSig) ,
\end{gathered}
\]
where
\[
\mathbf{L}_i = \left[ 
\begin{array}{ccc}
1 & 0 & w_i \\
0 & 1 & 0
\end{array}
\right] \text{ and } \bSig = \text{diag}(\sigma^2_1, \sigma^2_2).
\]
In terms of the CIHM specification (\ref{eq:full.model}), $\bp = (\bd, \sigma^2_1, \sigma^2_2)$. The parameters $\bg$ and $\bn$ are not used.

\subsection{Lapwing demographic modeling (IPM)}

The typical IPM \citep{schaub2011integrated} has the CIHM form
\[
[\by] = \prod_{i=1}^n[\by_i|\bg_i, \bn], \text{ and } [\bg] = \prod_{i=1}^n[\bg_i],
\]
where $\bg_i$ are data-type distinct parameters and $\bn$ are common parameters. Traditionally, there are no random unit-level effects ($\bt_i$) in IPM models. Multiple data sets are used to obtain better estimates of the common parameters, $\bn$, than can be obtained from any single $\by_i$. 

The lapwing data consist of $n=2$ data sets, a ring-recovery data set, $\by_1$ and an index measure of the abundance of adult female birds obtained from a generalized additive model of counts, $\by_2$. These data have been analyzed previously to demonstrate IPM methodology for ecological data \citep{besbeas2002integrating, besbeas2019exact, Brooks:2004zi, goudie2019joining}. 

Ring-recovery models are based on marking animals and releasing them back into the wild. In subsequent years, the marked animals are then recovered after they have died. The likelihood for this type of model is a product-multinomial where numbers of ring-recoveries in years following following release are multinomial distributed with cell probabilities that are functions of $\phi_t$ (for $t=1963,\dots,1997$), the probability of survival from time $t$ to $t+1$ and $\lambda_t$, the probability that an animal that died between $t-1$ and $t$ is recovered. To assess the influence of weather on survival and trends in recovery, the ring-recovery parameters are modeled with 
\begin{equation}
\begin{gathered}
\text{logit}(\phi_{yt}) = \eta_{y0} + \eta_{y1} x_t,\qquad \text{logit}(\phi_{at}) = \eta_{a0}+ \eta_{a1} x_t,\\
\text{logit}(\lambda_t) = \gamma_{\lambda 0} + \gamma_{\lambda 1} t,
\end{gathered}
\end{equation}
where $\phi_{yt}$ is the survival of first year birds, $\phi_{at}$ is adult female survival, and $x_t$ is the number of days below freezing. 

The census index data consist of noisy measures of female adult lapwing abundance in the study area. To make inference about population dynamics we use the state-space model of \citet{besbeas2002integrating},
\begin{equation}
\begin{gathered}
\mathbf{N}_{t+1} = \mathbf{T}_t\mathbf{N}_{t} + \boldsymbol{\epsilon}_{t}\\
y_{t} \sim \tN(N_{at}, \sigma^2),
\end{gathered}
\end{equation}

where $\mathbf{N}_t = (N_{yt}, N_{at})'$ is the number of yearling and adults females respectively in year $t$, $y_t$ are the noisy count of adult females. The transition matrix, $\mathbf{T}_t$, is parameterized with $\phi_t$ and the rate of offspring production, $\rho_t$. The survival parameters are the same as the ring-recovery model, but the production is modeled on the log scale using
\[
\log (\rho_t) = \gamma_{\rho 0} + \gamma_{\rho 1} t.
\]
Full model details are provided in online appendix S2. 

The IPM melds the information in the two data sets to allow inference for all parameters of the joint model, 
\[
\begin{gathered} 
\bn = (\eta_{y0}, \eta_{y1}, \eta_{a0}, \eta_{a1})',  \\
\bg_1 = (\gamma_{\lambda 0}, \gamma_{\lambda 1})' \text{, and } \bg_2 = (\gamma_{\rho 0}, \gamma_{\rho 1}, \log\sigma, \log N_{y0}, \log N_{a0})'.
\end{gathered}
\]
Random unit-level $\bt$ parameters are not traditionally used, but see the analysis in Section \ref{sec:analysis.ipm} for an alternate version. In each of the previous studies that analyze these data, bespoke code was written in software such as {\tt Matlab}, {\tt JAGS} \citep{plummer2003jags}, {\tt NIMBLE} \citep{de2017programming} or some other language to optimize the joint likelihood (integrated over latent abundance) or execute an MCMC. For a rather simple IPM model like this, it is straightforward to code. But for more complicated capture-recapture models or abundance  models this can become an impediment for practitioners. Many of the individual submodels in IPMs have software available for fitting them separately. For the ring-recovery data in this analysis there has been production-level software to fit the model for over 20 years. Thus, we should be able to utilize this substantial development in capture-recapture data analysis.

\section{Methods}

We propose a 2-stage method (extended to an optional 3rd stage) for analyzing CIHM models (\ref{eq:full.model}). In the first stage each data set, $\by_i$ is analyzed individually according to its individual data-level model in the CIHM using the best method and software available for that particular data type. Then, in the second stage, the results are combined (melded) using the unit-level and population-level models to approximate the inference as if the entire CIHM was fit in one step (e.g., via MCMC). In an optional third stage we can revisit estimates of the distinct unit parameters, $\bg_i$ to see if further updating is possible. 

\subsection{Two-stage approximate Bayesian inference} 

If we were to fit the CIHM model in the traditional way we would use summaries from the full posterior distribution 
\[
[\bp, \bn, \bg, \bt|\by] \propto [\by|\bt, \bg, \bn] [\bt|\bp][\bg][\bp,\bn].
\]
Suppose, however, we only possess a single $\by_i$. Then we might estimate $\bvt_i = (\bt_i', \bg_i', \bn')'$ using the posterior distribution
\[
[\bvt_i|\by_i] \propto [\by_i|\bvt_i] [\bvt_i],
\]
where $[\bvt_i]$ is a prior distribution that would be used in the absence of additional individuals. Using these individual posterior distributions, we can rewrite the full hierarchical model posterior as 
\begin{equation}
\label{eq:post}
[\bp, \bn, \bg, \bt|\by] \propto \left(\prod_{i=1}^n\frac{[\bvt_i|\by_i]}{[\bvt_i]}\right) [\bt|\bp][\bg][\bp,\bn].
\end{equation}


Stage I of the method we propose begins by first finding estimates $\hat{\bvt}_i$ and covariance matrices $\hat{\bS}_i$ such that we can approximate  
\begin{equation}
\label{eq:norm.approx}
\frac{[\bvt_i|\by_i]}{[\bvt_i]} = [\by_i|\bvt_i] \approx \text{N}(\bvt_i|\hat{\bvt}_i, \hat{\bS}_i)
\end{equation}
for each data set or partition. This approximation is motivated the standard large-sample asymptotic results. For large-sample theory to motivate the approximation, $[\by_i|\bvt_i]$ should satisfy the standard regularity conditions on the likelihood function, namely the ``true'' $\bvt_i$ is not on the edge of the parameter space and $[\by_i|\bvt_i]$ is continuously twice differentiable \citep{kass1989approximate, le2000asymptotics}. However, if the log-likelihood is approximately quadratic in $\bvt_i$, then the approximation can work well even for modest sample sizes \citep{geyer2005cam}. 

An estimate and covariance matrix for use in (\ref{eq:norm.approx}) is the maximum likelihood estimate (MLE) of $\bvt_i$ and $\hat{\bS}_i$ is the negative inverse Hessian. For now we assume that the size of $\by_i$ is large enough and the parameters identifiable enough such that $\hat{\bvt}_i$ is well defined and $\hat{\bS}_i^{-1}$ is nonnegative definite. The main benefit of using the MLE is that there is virtually always software available for finding the MLE and large-sample covariance matrix for common data-level models. 

If $\bvt_i$ is only weakly identified in $[\by_i|\bvt_i]$ then the MLE may be difficult to calculate. Thus, calculating estimate from the posterior distribution $[\bvt_i|\by_i] = [\by_i|\bvt_i][\bvt_i]$ might be desirable, where the temporary prior $[\bvt_i]$ is chosen to make all parameters identifiable. In this case, one might choose to use the maximum {\it a posteriori} (MAP) by optimizing $[\bvt_i|\by_i]$ and using the inverse Hessian for uncertainty as with the MLE. Or one could use a posterior sample (e.g., MCMC or importance sampling) to find the posterior mean and covariance matrix. Similar general regularity conditions for apply to the MLE, MAP, or MCMC estimates with respect to asymptotic approximation of well identified parameters. However, for nearly unidentifiable parameters in $[\bvt_i|\by_i]$, asymptotic approximations can be poor for even large samples because no matter how large the data set is, the posterior will be nearly (or exactly) equal to the prior distribution. Therefore, because we are free to choose the temporary prior to aid stage I model fitting, we suggest using a normal prior
\[
[\bvt_i] = \text{N}(\bvt_i|\bvt_0, \bS_0).
\]
A reparameterization of $\bvt_i$ might be necessary such that a normal density is appropriate as a prior, but this will help the accuracy of the normal approximation in stage I even for poorly identified parameters. Note that the support of the temporary prior needs to be the same as the desired parameter model $[\bt_i|\bp][\bg_i][\bn]$. With the use of the temporary prior, the approximation becomes,
\begin{equation}
\label{eq:pseudo.prior}
\frac{[\bvt_i|\by_i]}{[\bvt_i]} \approx \frac{\text{N}(\bvt_i|\Check{\bvt}_i, \Check{\bS}_i)}{\text{N}(\bvt_i|\bvt_0, \bS_0)} \propto \text{N}(\bvt_i|\hat{\bvt}_i, \hat{\bS}_i),
\end{equation}
where $\Check{\bvt}_i$ and $\Check{\bS}_i$ are the posterior mode (or mean) and  negative inverse Hessian (or covariance matrix) of $[\bvt_i|\by_i]$ and 
\begin{equation}
\label{eq:pprior.remove}
\hat{\bS}_i = \left(\Check{\bS}_i^{-1} - \bS_0^{-1}\right)^{-1} 
\ \ \text{and}\ \ 
\hat{\bvt}_i = \hat{\bS}_i \left(  \Check{\bS}_i^{-1}\Check{\bvt}_i -  \bS_0^{-1}\bvt_0\right). 
\end{equation}
This ratio was also used by \cite{goudie2019joining} in an approximate 2-stage MCMC method. 

Using (\ref{eq:post}) and (\ref{eq:norm.approx}) for $[\bvt_i|\by_i]/[\bvt_i]$, the desired full posterior density function for all the parameters can be approximated (up to a proportional constant) by
\begin{equation}
\label{eq:normpost}
[\bp,\bn, \bg, \bt|\by] \approx  \left(\prod_{i=1}^n\text{N}(\hat{\bvt}_i|\bvt_i, \hat{\bS}_i)\right) [\bt|\bp][\bg] [\bp,\bn].
\end{equation}
Note, in (\ref{eq:normpost}) we have put $\bvt_i$ in the mean position of the normal specification. Due to the symmetry of the normal density function it is equivalent, but this way $\hat{\bvt}_i$ can be interpreted as an observation. 
In stage II we can make full inference for all model parameters by maximizing the approximate full posterior (\ref{eq:normpost}) to obtain the posterior mode $(\tilde{\bp}',\tilde{\bn}',\tilde{\bg}',\tilde{\bt})'$, or {\it maximum a posteriori} (MAP) estimate and the negative inverse Hessian of the log posterior density estimates the posterior covariance matrix. Hats are used to represent stage I estimates and tildes to represent stage II estimates. For the remainder of the paper we will assume that some measure of location, $\hat{\bvt}_i$, and scale $\hat{\bS}_i$ are available where $\hat{\bS}_i^{-1}$ is nonnegative definite. These can come from any combination of MLEs, MAPs, or MCMC samples, whatever software is most appropriate or available for the data model can be used. Different methods can be used between stage I units as well. 

\subsection{Normal random unit-level effects}

While there is no requirement that the unit-level random effects, $\bt$, be normally distributed in (\ref{eq:normpost}), it is the case in the vast majority of applications, or $\bt_i$ can often be parameterized such that a normal distribution is appropriate. Thus, for the remainder of the paper, we assume that, 
\[
[\bt|\bp] = \text{N}(\bt|\bm, \bSig),
\]
where $\bp = (\bm', \bs')'$ and $\bSig$ is a covariance matrix parameterized by $\bs$. When this is the case, stage II becomes equivalent to fitting a multivariate normal random effects linear model. 

First we reparameterize the model using $\bt_i = \bm + \bu_i$ where $\bu = (\bu_1',\dots,\bu_n')'\sim \tN(\mathbf{0},\bSig)$, $\ba = (\bn', \bg_1',\dots,\bg_n')'$, $\ba_i = (\bn', \bg_i')'$, and $\bb=(\bm', \ba')'$. Now, the approximation (\ref{eq:normpost}) becomes
\begin{equation}
\label{eq:normeff}
[\bm,\bs, \bn, \bg, \bu|\by] \approx  \prod_{i=1}^n\left(\tN(\hat{\bvt}_i|\bX_i\bb+\bZ_i\bu_i, \hat{\bS}_i)\right) \tN(\bu|\mathbf{0},\bSig) [\bg][\bm, \bs][\bn],
\end{equation}
where $\bX_i = \text{diag}(\bI, \bK_i)$, $\bI$ is a $|\bm|$ size identity matrix, $\bK_i$ is a $|\ba_i|\times |\ba|$ indicator matrix that selects the appropriate parameter, and $\bZ_i$ is a $|\bvt_i|\times |\bt_i|$ indicator matrix that selects $\bu_i$ to align with the $\bt_i$ portion of $\bvt_i$. If $\bm$ is represented as a linear model as in the moose browse example, then $\bI$ is replaced with the appropriate design matrix (e.g., $\bL_i$ in Section \ref{sec:moose.data}). After concatenation, $\bvt = (\bvt_1',\dots,\bvt_n')'$, $\bX = (\bX_1',\dots,\bX_n')'$, $\bZ = (\bZ_1',\dots,\bZ_n')'$ and $\boldsymbol{\epsilon} \sim \tN(\mathbf{0}, \text{diag}(\hat{\bS}_1,\dots,\hat{\bS}_n))$, finding the MAP of $(\bm,\bs, \bn, \bg, \bu)$ is equivalent to fitting the multivariate linear mixed model
\begin{equation}\label{eq:full2stage}
\hat{\bvt} = \bX\bb + \bZ\bu + \boldsymbol{\epsilon},
\end{equation}
subject to the prior distribution $[\bg][\bm, \bs][\bn]$.     

In the analysis of the two ecological data sets, we use the {\tt R} package {\tt TMB} to fit (\ref{eq:normeff}), but it is just a multivariate linear mixed model, so any software that can account for the parameter prior distributions can be used. {\tt TMB} uses Laplace's method to find the MAP of $\bp = (\bm,\bs)$, $\bn$, and $\bg$ using the marginal posterior $[\bm,\bs, \bn, \bg|\by]$ then a generalized delta-method \citep{kass1989approximate} is employed to estimate $\bu$ accounting for uncertainty in the other parameters \citep{kristensen2016tmb,skaug2006automatic}. At first it might seem as if we are adding an another layer of approximation by using the Laplace method, however, given the normal approximation already made in (\ref{eq:normpost}) and the normality of the unit-level parameters, the Laplace method of approximating the marginal posterior density of $(\bm,\bs, \bn, \bg)|\by$ obtained from (\ref{eq:normeff}) is actually exact because the log-posterior is quadratic in $\bu$ \citep{goutis1999explaining}. After finding the MAP, traditionally, the inverse Hessian of (\ref{eq:normeff}) is used to estimate the covariance matrix of the full model (\ref{eq:normeff}), which is what we use in the examples that follow.

\subsection{Stage III: revisiting stage I}  

Using the description to this point, we are set to perform 2-stage inference for a CIHM. However, we noticed, in our initial analysis of the lapwing data in forthcoming Section 4.2, when $\bS_i^{-1} \approx \mathbf{0}$ (relative to $\bvt_i$) for a particular unit in in stage I, then updating $\hat{\bg}_i$ might not be as efficient as desired. Therefore, we propose revisiting stage I estimation to update $\bg_i$ given the results of the second stage using the generalized delta method initially proposed by \citep{kass1989approximate}. Note, that we are not proposing that stage III is always necessary, but, that it is an option, especially if $\bg_i$ is not well identified in the data model alone. 

For a heuristic example of lack of updating for $\hat\bg_i$ consider the simple 2 parameter model structure:
\begin{equation}
\by_1 \sim [\by_1|\eta]\ \text{ and } \by_2 \sim [\by_2|\eta,\gamma_2].
\end{equation} 
Further, suppose that $\eta$ and $\gamma_2$ jointly are only vaguely identifiable in $[\by_2|\eta,\gamma_2]$, whereas it is fully identifiable in $[\by_1|\eta]$. In stage I, we can obtain well informed estimates $\hat{\bn}_1$ and $\hat{\bS}_1$. However, for $\by_2$, $\hat{\bS}_2^{-1}$ will be small. In the extreme case that $\hat{\bS}_2^{-1} \to \bz$ one can show  (through the generalized least-square estimate of $\bb = (\bn,\bg_2)$ in (\ref{eq:full2stage})), that $\tilde{\bn} \to \hat{\bn}_1$ for any $\hat\bn_2$. However, $\tilde{\bg}_2 \to \hat{\bg}_2$, remaining unchanged from stage I. So, we see satisfactory updating of $\hat{\bn}_2$, but no updating of $\hat{\bg}_2$ in stage II. This case is admittedly pathological, but it illustrates that the Hessian matrix of $[\by_2|\bn=\hat{\bn}_2, \hat{\bg}_2]$ may not be the best representation of the likelihood shape at the full MAP. 

Stage III proceeds by first marginalizing over $\bg_i$. This can be done analytically if possible. However, based on the initial normal approximation after stage I, after $\hat{\bvt_i}$ and $\hat{\bS}_i$ are calculated, the rows and columns associated with $\bg_i$ can be ignored when moving to stage II. Let $\bxi$ represent all parameters except $\bg_i$, after stage II, we obtain the marginal posterior approximation:
\[
[\bxi|\bY] \approx \tN(\bxi\mid \tilde{\bxi}, \widetilde{\bV}_{\xi}),
\]  
where $\widetilde{\bV}_{\xi_i}$ is the negative inverse of the Hessian of the log-posterior distribution represented by the stage II model (\ref{eq:normeff}) and the associated prior distributions. If $\bg_i$ is a nuisance parameter and of little interest, we can stop here and there is no need for stage III. However, if as estimate of $\bg_i$ is desired, we can obtain an updated mean and covariance matrix for the full posterior approximation such that
\[
[\bg_i ,\bxi|\bY] \approx \tN((\bg_i ,\bxi) \mid \bar{\bxi}, \overline{\bV}),
\]
where $\bar{\bxi} = (\bar{\bg}_i,\tilde{\bxi})$, $\bar{\bg}_i$ is the value that maximizes the conditional posterior $[\bg_i|\tilde{\bt}_i, \tilde{\bn}, \by_i]$ and $\overline{\bV}$ is a updated posterior covariance matrix. To calculate $\overline{\bV}$, we need to define the matrices: 
\[
\begin{gathered}
\mathbf{H}_{\gamma\gamma} = \frac{\partial^2}{\partial \bg_i \partial \bg_i'} \left\{\log[\by_i|\tilde{\bt}_i, \tilde{\bn}, \bar{\bg}_i] + \log[\bar{\bg}_i] \right\}
\text{ and }
\mathbf{H}_{\gamma\xi} = \frac{\partial^2}{\partial \bg_i \partial \bxi_i'} \log[\by_i|\tilde{\bt}_i, \tilde{\bn}, \bar{\bg}_i].
\end{gathered}
\] 
Now the full covariance matrix can be approximated by \citep{kass1989approximate, skaug2006automatic}:
\[
\begin{gathered}
\overline{\bV} = \left[\begin{array}{cc} -\mathbf{H}_{\gamma\gamma}^{-1} & \bz \\ \bz & \bz \end{array}\right] + 
\mathbf{J} \widetilde{\bV}_{\xi} \mathbf{J}',  \text{ where } 
\mathbf{J} = \left[ \begin{array}{c} -\mathbf{H}_{\gamma\gamma}^{-1}\mathbf{H}_{\gamma\xi} \\ \mathbf{I}_{\xi} \end{array}\right].
\end{gathered}
\]
This may not be applicable in all situations, because one must be able to evaluate $[\bg_i|\tilde{\bt}_i, \tilde{\bn}, \by_i]$ to a proportional constant as well as its gradient and Hessian for various parameter values. This may not be practical for most production software used in stage I. 

One could question whether or not $\bt_i$ might suffer the same lack of sufficient updating in stage II as well as $\bg_i$ if the likelihood is generally flat. However, this is much less likely due to the fact that there is a unit-level model, $[\bt|\bp]$, which is being explicitly fit in the second stage. This allows information in other data sets to directly influence and update $\bt_i$ at stage II. However, $\by_j$, $j\ne i$, can only contribute to updates of $\hat\bg_i$ via the off-diagonal elements of $\hat\bS_i^{-1}$.

\section{Model fitting and results} \label{sec:analysis}

\subsection{Moose browse GLMM}    

To effectively use the 2-stage approach in this example, it was necessary to use a stage I temporary prior (\ref{eq:pseudo.prior}) and (\ref{eq:pprior.remove}) because browsing is rare (or absent) in some of the plantations. We used the normal $g$-prior \citep{hanson2014informative} where $\bt_0 = \mathbf{0}$, $\bS_0 = \pi^2 n_i (\mathbf{H}_i'\mathbf{H}_i)^{-1}/6$, and $\mathbf{H}_i$ is the usual design matrix with a column for intercept and the other containing tree height ($h_{ij}$). This prior produces $p_{ij}$ that are approximately uniform(0,1) distributed for a randomly selected  $h_{ij}$. The effect of this prior was removed prior to stage II using (\ref{eq:pprior.remove}).

For stage I model fitting we used two approaches. In the first version (A), we fitted each individual model using the {\tt R} package {\tt mgcv} \citep{wood2011fast} to obtain the posterior mode and Hessian covariance matrix from $[\bt_i|\by_i] \propto [\by_i|\bt_i][\bt_i]$. In the second version (B), we used the approximate posterior mean and covariance matrix obtained from a deterministic posterior sampling procedure detailed in \citet{johnson2011bayesian} (see online appendix S1). This deterministic procedure estimates the posterior mean and variance without asymptotic results. For version B, the likelihood must be calculated with bespoke code. We are not aware of any production-level software that returns only the value of the posterior for logistic regression, which makes this version more challenging in practice than the first version, but we explored here nonetheless. In stage II we used {\tt TMB} \citep{kristensen2016tmb} to fit the linear mixed model in (\ref{eq:full2stage}). Code for this example as well as the IPM example are provided in an online supplement. 

\begin{figure}
\includegraphics[width=6.5in]{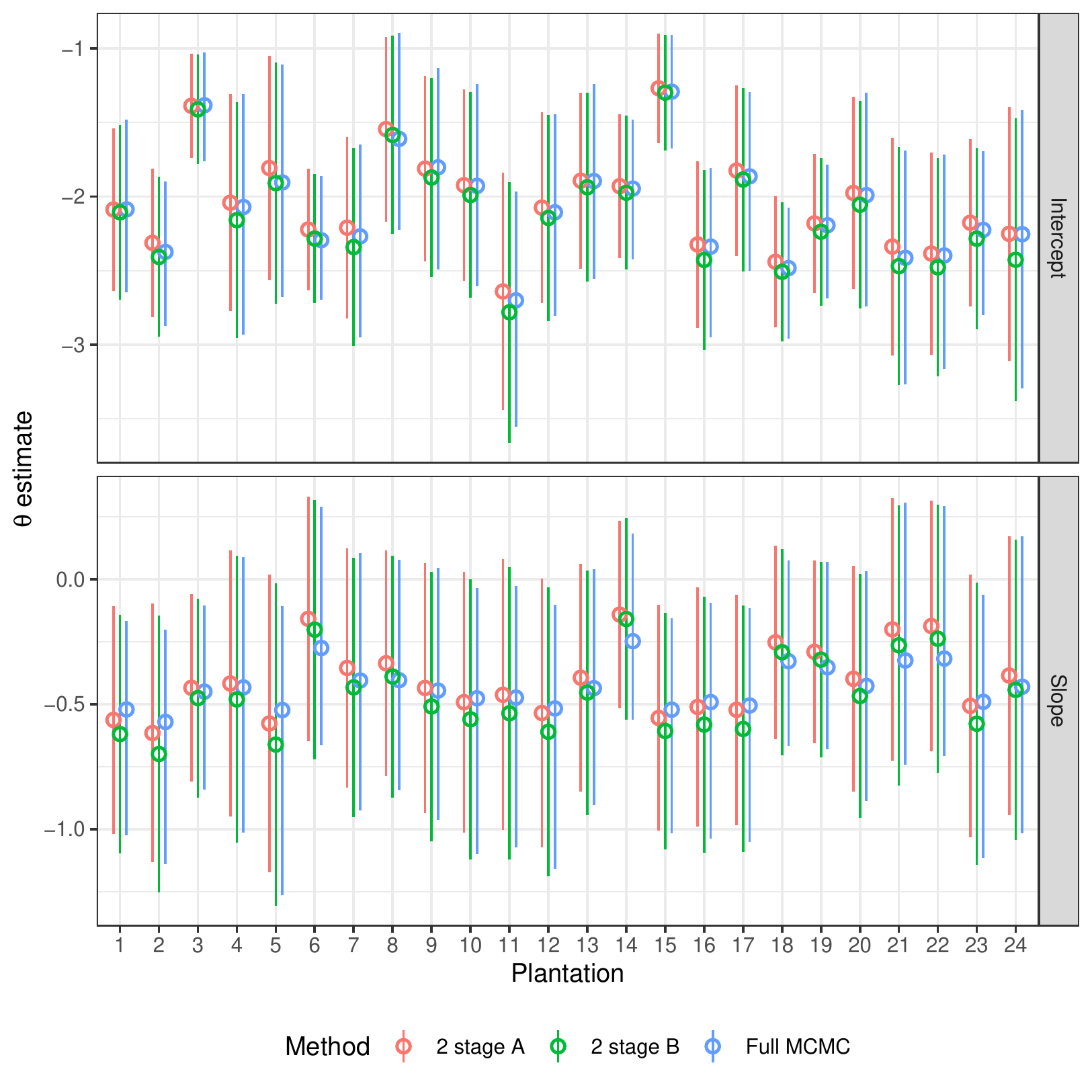}
\caption{\label{fig:theta.re}Random effect inference for Example I using 2-stage and full MCMC methods. In the 2-stage analyses this was estimated using $\tilde{\bt}_i = \bX_i\tilde{\bb} + \tilde{u}_i$, where the tilde represents second stage estimates.}
\end{figure}

\begin{figure}
\includegraphics[width=6.5in]{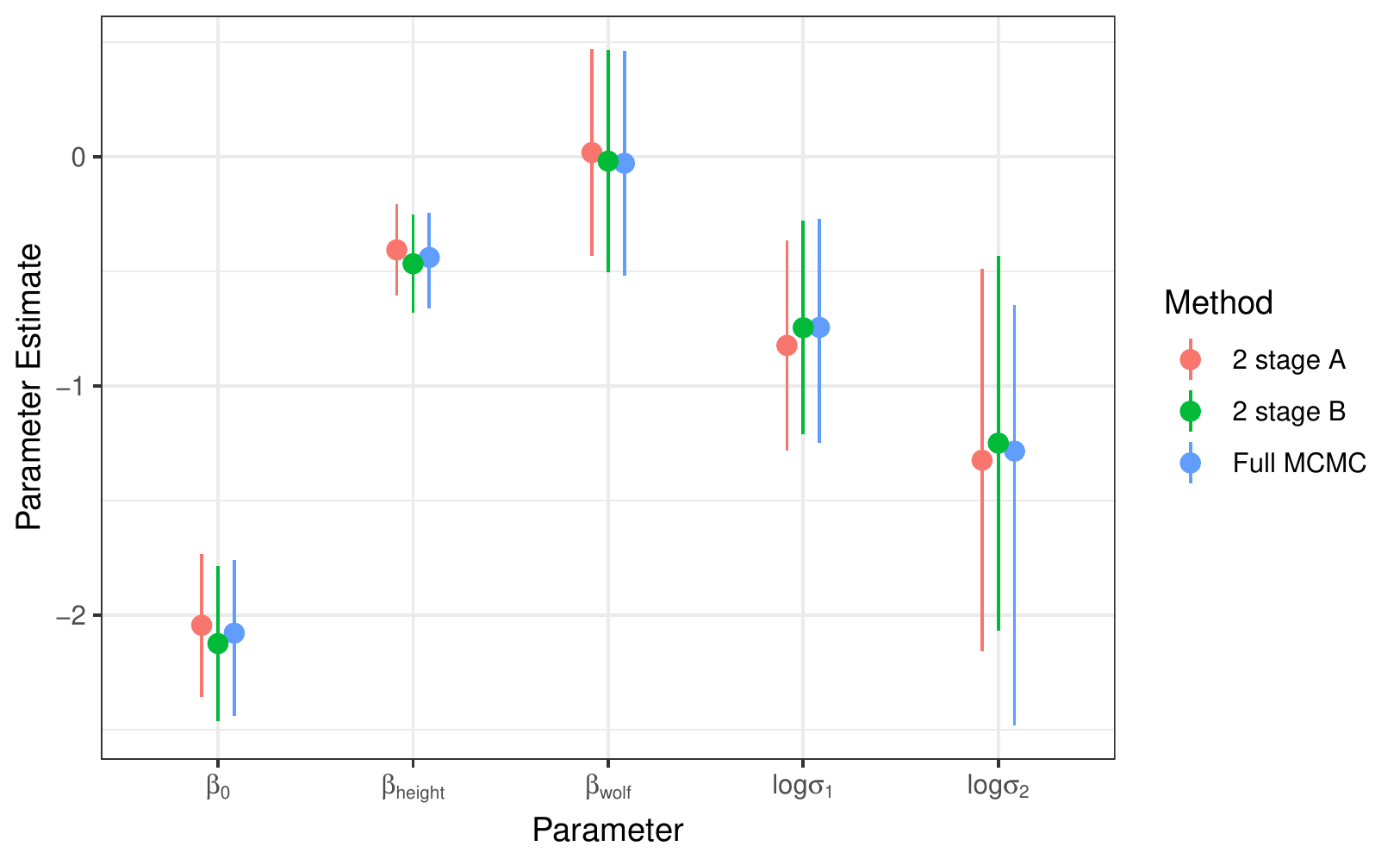}
\caption{\label{fig:fixed.re}Fixed effect and variance parameter inference for Example I using 2-stage and full MCMC methods.}
\end{figure}

We examined predictive estimates $\tilde{\bt}_i = \bX_i\tilde{\bb} + \tilde{\bu}_i$, fixed effects $\tilde{\bb}$, and variance parameters $(\log\tilde{\sigma}_1, \log\tilde{\sigma}_2)$ with respect to a full MCMC analysis. Both versions of the 2-stage approach produced results similar to a gold standard MCMC analysis using all the data simultaneously (Figures \ref{fig:theta.re} and \ref{fig:fixed.re}). Heuristically, for both $\bt_i$ and $\bSig$,  the mean and covariance matrix obtained from stage I, version B (deterministic posterior sample), produced estimates more similar to the full-model MCMC estimates than those derived from the Bayesian MAP fitting with {\tt mgcv}, but only slightly so. Scientific inference for any of the three methods would be identical, especially for the fixed parameters $\bb$.

\subsection{Lapwing IPM} \label{sec:analysis.ipm}

To demonstrate how we might utilize production software, in the first stage, we fitted the ring-recovery model using {\tt MARK} \citep{white1999program} via the user interface {\tt R} package {\tt RMark} \citep{laake2013:RMark}. {\tt MARK} provides the MLE and the large-sample covariance matrix for $\hat{\ba}_1 = (\hat{\bn}_1, \hat{\bg}_1)'$ and $\hat{\bS}_1$.  For the state-space model of the female count data, we used bespoke model code written for analysis with the {\tt R} package {\tt TMB} \citep{kristensen2016tmb}. However, the state-space model is the simpler of the two data models in terms of programming complexity. From {\tt TMB} we obtain the MAP estimate and large-sample covariance matrix, $\hat{\ba}_2 = (\hat{\bn}_2, \hat{\bg}_2)'$ and $\hat{\bS}_2$. Because the survival and production parameters are not well identified in the state-space model, we used vaguely informative temporary stage I prior distributions. For $(\eta_{y0},\eta_{y1})$ and $(\eta_{a0},\eta_{a1})$ we used independent normal $g$-priors to induce approximately uniform priors on $\phi_{yt}$ and $\phi_{at}$ \citep{hanson2014informative}. For the production parameters, $(\gamma_{0\rho},\gamma_{1\rho})$ we used a normal $g$-prior such that $\rho_t$ lies within [0.01, 2] with prior probability $\approx 0.95$. Based on other analyses, this prior covers a fairly broad range of values for the average number of females produced per female. For the remaining parameters, temporary priors were not used.  In stage II, we used the linear mixed model in (\ref{eq:full2stage}) along with the population prior distributions:
\[
\begin{gathered}[]
[\eta_{y0}, \eta_{y1}, \eta_{a0}, \eta_{a1}] = \tN(\mathbf{0}, 10^2\bI),\\
[\gamma_{\lambda 0}, \gamma_{\lambda 1}] = \tN(\mathbf{0}, 10^2\bI),\ [\gamma_{\rho 0}, \gamma_{\rho 1}] = \tN(\mathbf{0}, 10^2\bI) \\
[\log N_{y0}, \log N_{a0}] = \tN(0, 1000^2\bI), \text{ and }[\log \sigma] \propto \text{constant (improper)}.
\end{gathered}
\]

We also consider an additional complexity beyond the traditional IPM specification. One of the tenets of the IPM is that the multiple data models share common parameters of scientific interest; in our case, the survival parameters, $\bn$. It seems reasonable, however, that differences in data collection procedures or model specification might imply separate ``true'' values of the survival parameters, that is, $\hat{\bn}_i$ is actually estimating $\bn_i\ne\bn$. So, one might consider each $[\by_i|\bn,\bg_i]$ to be an imperfect source of information about $\bn$. This is how meta-analysts would traditionally envision combining these separate studies. Therefore, we also fit a model with random effects $\bt_i \sim \tN(\bn, \sigma^2_\theta \bI)$.
If the available data suggest common survival parameters between the two data sets, then $\tilde{\sigma}_\theta$ will be small and all the stage II estimates of each $\tilde{\bt}_i \approx \tilde{\bn}$. However, if $\tilde\sigma_\theta \gg 0$, then a fixed $\bn$ model is probably inappropriate and there are possible biases in one or both of the different data sets. Thus, $\sigma_\theta$ serves as a regulator to assess the degree to which the data sets support a fixed $\bn$ value. For this model we used $[\sigma_\theta]$ = Exponential(1) in addition to the previous prior specifications. 

\begin{figure}
\includegraphics[width=6.5in]{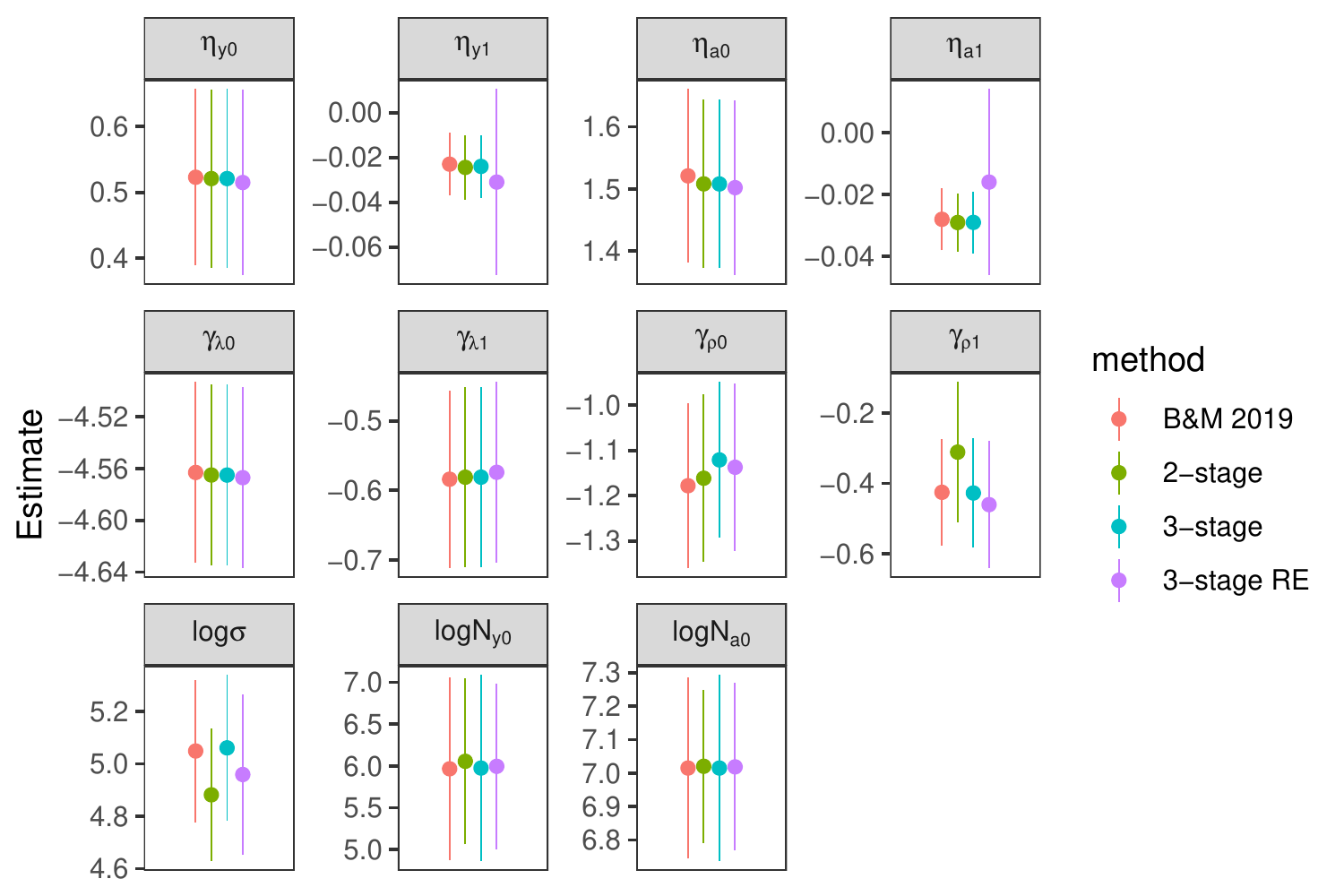} 
\caption{\label{fig:ipm} Parameter estimates for the lapwing IPM model. The B\&M 2019 results were taken from Table 2 of \citet{besbeas2019exact} and represent estimates from single stage estimation of the full model. In the 3-stage RE model, $\bt_i \sim \tN(\bn,\sigma_{\theta}^2\bI)$. Parameter estimates are also provided in online supplement Table S1.}
\end{figure}

To judge the quality of the posterior approximation, we compared our results to those presented by \citet{besbeas2019exact}, hereafter, B\&M. Table 2 (pg. 483) of that publication provides the results of a full joint analysis for several different models and inference methods. Our specification is most similar to their ``MLE KF'' model. That is a model where $\log N_{y0}$ and $\log N_{a0}$ are explicitly estimated.  We only compare our results to that model (Figure \ref{fig:ipm}), but the other results are very similar. All of the multistage approximate fitting methods produced results very similar to  the B\&M results (Figure \ref{fig:ipm}). A full table of the multistage results is available in Appendix B. For the 2 and 3 stage fixed $\bn$ models, estimates of $\bn$ were virtually indistinguishable from those given by B\&M, both point estimates and standard deviations. Also, the band reporting parameters, $\bg_1$, are nearly identical. From stage II to stage III, $\bar{\gamma}_{\rho 1}$ became more similar to the B\&M results while $\bar{\gamma}_{\rho 0}$ became less similar. Although they are both close to the B\&M results overall. A marked improvement in estimates from stage II to III is evident in the $\log \sigma$, $\log N_{y0}$, and $\log N_{a0}$ parameters, with all of them becoming nearly equivalent to the B\&M estimates after stage III.  

\begin{figure}
\includegraphics[width=6.5in]{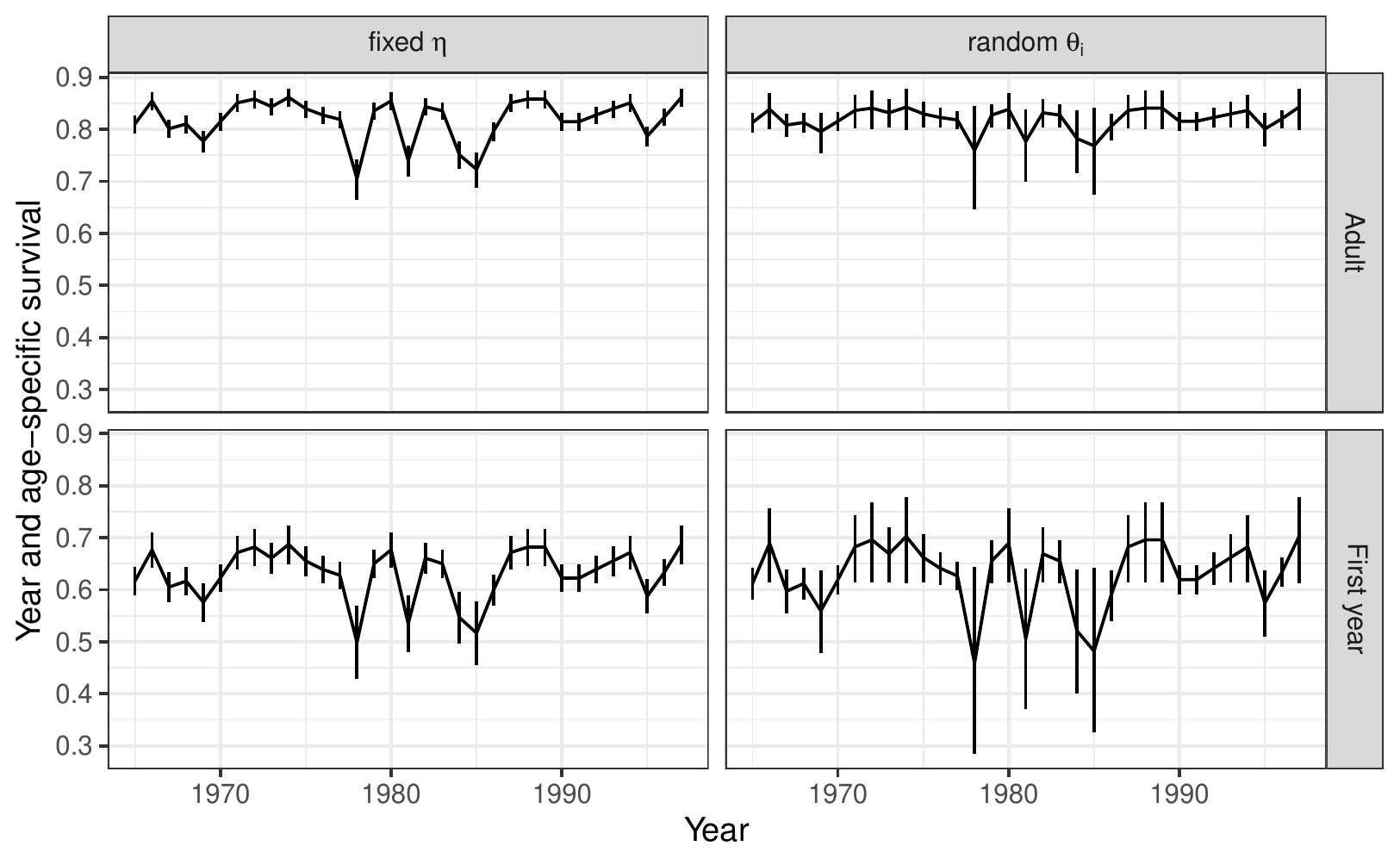} 
\caption{\label{fig:surv}Estimates of year and age-specific survival for lapwings. The different age classes are represented in the rows; adult and first year. The columns illustrate estimates from the two different models. The first column is the traditional IPM where the survival parameters are common between the data sets (i.e., $\bn$ is fixed). In the second column we modeled $\bt_i \sim \tN(\bn, \sigma_{\theta}^2\bI)$, where $\sigma_{\theta}$ is estimated in the second stage.}
\end{figure}

We now turn our attention to the random effect version of the model. The estimate of $\tilde{\sigma}_\eta$ = 0.017 (0.006--0.049), which is small, but not trivial. Thus, there is some evidence that each data set contains different information about the true survival process. The population-level effect of frost days on survival becomes slightly steeper and more uncertain (Figure \ref{fig:ipm}) for young animals. For adult animals, frost days have less effect on survival, but this is again, more uncertain. The increase in uncertainty and change in frost effect are plainly evident when examining survival on the natural scale (Figure \ref{fig:surv}). Although they did not fit separate survival processes, \citet{goudie2019joining} noted this discrepancy as well in their 2-stage MCMC analysis of these same data.

\section{Discussion}

We have proposed a multistage method for estimating parameters in a broad class of BHMs. Our approach is computationally efficient, flexible, and draws inspiration from traditional meta-analysis where individual studies using differing methods contribute unit-level parameter estimates which are then combined using a CIHM format. In stage I, practitioners can use software expressly designed for a particular data model. We use traditional GLM/GAM software in the first example, and a mixture of production software ({\tt MARK}/{\tt RMark}) and pseudo-coding software ({\tt TMB}) in the IPM example. This increases the efficiency of the analysis in development time because software expressly designed for the data at-hand is being used. Analysis do not have to create bespoke MCMC that will often be substandard to production level software in terms of ease-of-use and computational speed. In addition, as with other 2-stage methods, the stage I analyses are ``embarrassingly parallel,'' so, they can be executed in parallel if necessary, further increasing efficiency. In stage II, the multivariate linear mixed model is very quick to analyze because the dimension of the problem is substantially reduced from the data models. The linear mixed model represents a compression of the information in the various data sets about the parameters. Finally, the generalized delta-method allows one to revisit the estimates of distinct parameters that is usually not possible in traditional meta-analysis because the raw data are not available.  

This efficiency comes at a cost, however, because it relies on approximations in two places during the development, the normal approximation in stage I and the normal approximation of the stage II posterior distribution. We do not view this as a detriment because this method is designed for computationally challenging situations. These usually occur in settings with a large amount of data and hence a substantial amount of information about the parameters. That is, situations for which the large sample approximations were developed. Both of the models in our examples could be fitted readily in a single stage BHM analysis. Even so, the multistage method produced estimates very close to the ``gold standard'' full BHM estimates. The 2-stage approach may not be optimal in all settings, thus if one is capable of fitting the full CIHM in one step that would be the recommended course of action. We suggest our 2-stage approach for instances where computation or development of software to fit the full model at once is impractical. 

Even though the approximations produced good results in our examples we should examine the seriousness of these approximations and what options practitioners have for checking the degree of error introduced by these approximations. The first approximation is the Gaussian approximation of the data-level likelihood shape at the first stage. This approximation is justified by the standard large sample results that coincide for MLE and posterior modes (MAP) and means. There are some structural situations with stage I models where the assumptions underlying those results would be questionable. \citet[][Section 4.2]{gelman2013bayesian} presents several scenarios where the assumptions of large sample theory fail to be met. Most of these situations are able to be fixed with appropriate reparameterization or prior selection. For example, parameters should be specified on the log or logit scale, as well as the use of the Gaussian temporary prior in (\ref{eq:pseudo.prior}). 

If the data-level models are sufficiently regular, one may still want to assess whether the sample size is sufficient to warrant a Gaussian approximation. We note, in a variety of previous studies this assumption has often worked well with little technical justification (e.g., \citealt{geyer2005cam}, \citealt{goudie2019joining}, \citealt{scott2016bayes}). If the validity of the MLE or MAP estimates and Hessian derived covariance are in question, one can also use bootstrap bias and variance corrections to improve the results \citep{geyer2005cam,scott2016bayes}. One can also increase data size by combining the original data partitions where appropriate. To ascertain whether sample size is too small for the shape of the normal density function to be appropriate one could also perform a MCMC analysis of some of the smallest units to see how posterior density appears. If the approximating normal posterior is used as a proposal distribution in a simple independent Metropolis-Hastings algorithm, then the acceptance rate can be used to judge the closeness of the approximation. For example, an acceptance rate of 0.9 (i.e., 90\% of proposals accepted) would imply that the stage I approximation is very close to the actual stage I posterior shape. \citet{michelot2018langevin} use this type of diagnostic for assessing adequacy of Euler approximations for Langevin diffusion models. Because the partitions were designed such that individual data models are relatively easy to evaluate this should pose little challenge for at least a few low sample size partitions. After diagnostic checking, if a normal approximation appears  inappropriate, \citet{gelman2013bayesian} suggests a normal mixture,  
\[
[\by_i|\bt_i] \approx \sum_{k=1}^{K_i}\omega_{ik}\text{N}(\bvt_{i}|\bvt_{ik}^*, \bS_{ik}^*),
\]
where $\sum_k \omega_{ik} = 1$, and $\{\bvt_{ik}^*, \bS_{ik}^*\}$ are selected in an advantageous way in stage I. If $\bt_i \sim \text{N}(\bm, \bSig)$ then we can follow the same development in Section 3.2 to obtain the stage II model that is a mixture of linear mixed models. That is, the approximate posterior would be
\[
[\bm,\bs, \bn, \bg, \bu|\by] \propto  \prod_{i=1}^n\left(\sum_{k=1}^{K_i}\omega_{ik}\tN(\bvt_{ik}^*|\bX_i\bb+\bZ_i\bu_{ik}, \bS_{ik}^*)\right) \tN(\bu|\mathbf{0},\bSig) [\bg][\bm, \bs][\bn].
\]
Although it is no longer a traditional linear mixed model, it is still relatively easy to fit because the weights are known. 

In stage II, when using the generalized delta method for estimating $\bt$ and the stage III method for reassessing $\bg$, we make the additional assumption that $[\bm,\bs,\bn,\bg|\by]$ is approximately normal. Again, given we have already made the assumption that the stage I posteriors are approximately normal. This new approximation is not adding a substantial new element of error to the analysis because, for a given value of the random effect variance parameters, $\bs$, normal random effects, and the original normal assumption, the stage II posterior is multivariate normal (assuming normal priors on $\bb$). However, if needed, then bootstrap, MCMC or other method could be used for the stage II inference instead of the MAP and Hessian. 

We developed the proposed multistage method under the assumptions of the CIHM (\ref{eq:full.model}), however, we  can relax some of the independence assumptions in the CIHM at the data level and the method is still valid in its current form. First, the data do not need to be strictly independent at the unit-level. That is, the multistage method can still be applied when the data models can be partitioned as
\[
\by \sim \prod_i[\by_i|\by_{-i}, \bt_i, \bg_i, \bn],
\]
where, $\by_{-i} = \{\by_1,\dots,\by_{i-1},\by_{i+1},\dots,\by_n\}$. Having a {\it meaningful} partition of the data is not the important part of any multistage analysis, it is the ability to partition the data-level model \citep{hooten2018prior}. In stage I, these dependent data models can be individually fit in the same way.

Hierarchical Bayesian models have brought a wealth of analytical capability for extracting scientific signal out of complex and noisy data. Moreover they provide an honest assessment of uncertainty by accounting for many different forms of stochasticity and relationships in the observable data. By assembling relatively simple components, a complex model can emerge. With the advent of modern data collection capabilities, default methods for inference have started to become obsolete. \citet{gelfand2015hierarchical} note that BHMs have been a great benefit to science but the current rate of data acquisition and model complexity are ``beginning to stretch our computational capabilities'' and future inference will have to be the result of ``an artful mixture of model specification and approximation.'' Hierarchical conditioning provides a beneficial way to build large complex models, thus it is sensible that inferential computing for these models may also be best accomplished in a hierarchical fashion \citep{mccaslin2021hierarchical}.

\section*{Acknowledgments}
The findings and conclusions of the NOAA authors in the paper are their own and do not necessarily represent the views of the National Marine Fisheries Service, NOAA.  Any use of trade, firm, or product names is for descriptive purposes only and does not imply endorsement by the U.S. Government.  Funding for this research was provided by NSF DEB 1927177.

\bibliography{bibLibrary.bib}

\clearpage

\clearpage

\clearpage

\clearpage

\clearpage

\end{document}


\renewcommand{\baselinestretch}{1.75}\normalsize
\raggedright
\setlength{\parindent}{2em}
\raggedbottom

\begin{center}
{\Large {\bf Online Supplement} \\ Greater Than the Sum of its Parts: \\Computationally Flexible Bayesian Hierarchical Modeling}
\end{center}

\bigskip

\linenumbers

\section*{S1. Deterministic Bayesian inference} 
\label{sec:methodii}

To this point we have really only considered using the posterior mode and Hessian derived covariance matrix for the first stage normal approximation. For a large amount of data in the first stage, this might be just fine. But, we might consider using the posterior mean and variance of $\bt_i$ instead. If MCMC is used in the first stage, we  calculate the sample mean and covariance matrix. If this is the case, one might use the two-stage MCMC procedures proposed and discussed by \cite{hooten2018prior}, \cite{lunn2013fully}, and \cite{goudie2019joining}. If the number of parameters is relatively small, say $\le 6$, we propose using a deterministic sampling procedure for approximating the posterior mean and variance (see \citealt{johnson2011bayesian}). 

The deterministic procedure proceeds as follows for each $i$,.
\begin{enumerate}
\item Maximize $[\bt|\by]$, to obtain the posterior mode, $\hat{\bt}$ and covariance matrix, $\hat{\bS}$
\item Form the Eigen decomposition of the covariance matrix $\hat{\bS} = \mathbf{V}\boldsymbol{\Lambda}\mathbf{V}'$
\item Explore the principle axes of $[\bt|\by]$ via the parameterization $\bt^{(j)} = \hat{\bt} + \mathbf{V}\boldsymbol{\Lambda}^{1/2}\mathbf{z}$, where $\mathbf{z}$ is successively incremented from $\mathbf{0}$ one entry at a time by step length, say $\delta_z$ until $\log[\hat{\bt}|\by]-\log[\bt^{(j)}|\by] > \delta_\pi$.
\item Repeat by successively incrementing each element of $\mathbf{z}$ by $-\delta_z$.
\item Finally, form a grid with every combination of the dimensionality $\bt^{(j)}$ entries saved in the previous two steps and retain with the same $\delta_\pi$ criterion.
\item Form weights $w_j \propto [\bt^{(j)}|\by]$
\end{enumerate}
The main benefit of the deterministic approach is that it avoids having to select a proposal distribution and assess chain convergence. The drawbacks of this approach, however, are the unknown number of likelihood evaluations and the potential coarseness in high density areas. As the number of parameters becomes large this method quickly succumbs to the curse of dimensionality.

\clearpage 

\section*{S2. Additional details and results for Example II}
\label{sec:ipm.results}

\subsection*{S2.1. Model Details}

Ring-recovery models are based on marking animals and releasing them back into the wild. In subsequent years, the marked animals are then recovered after they have died. In this study lapwings were released and recovered annually from 1963--1997. The likelihood for this type of model is a product-multinomial where numbers of ring-recoveries in the following years are multinomial distributed with cell probabilities
\[
p_{it} = \left\{ 
\begin{array}{ll}
(1-\phi_{t-1})\lambda_t & \text{ for } t=i+1 \\
(1-\phi_{t-1})\lambda_t\prod_{t'=I}^{t-2} \phi_{t'} & \text{ for } i+2 < t \le T \\
1-\sum_{t'=i}^J p_{it'} & \text{ for } t=T+1
\end{array} \right. ,
\]
where $p_{it}$ is the probability that an animal ringed in year $i$ is recovered in year $t$. The last year of the study is denoted with $T$, ($t=T+1$ is for animals never recovered), $\lambda_t$ is the probability of recovering an animal in year $t$ given it died over the previous year, and $\phi_t$ is the probability of surviving from year $t$ to $t+1$. 

To assess the influence of weather on survival and trends in recovery, the ring-recovery parameters are modeled with 
\begin{equation}
\begin{gathered}
\text{logit}(\phi_{yt}) = \eta_{y0} + \eta_{y1} x_t,\qquad \text{logit}(\phi_{at}) = \eta_{a0}+ \eta_{a1} x_t,\\
\text{logit}(\lambda_t) = \gamma_{\lambda 0} + \gamma_{\lambda 1} t,
\end{gathered}
\end{equation}
where $\phi_{yt}$ is the survival of first year birds in year $t$ (for animals ringed in year $t-1$), $\phi_{at}$ is adult female survival in year $t$, and $x_t$ is the number of days below freezing in year $t$. 

The census index data consist of noisy measures of female adult lapwing abundance in the study area, $\by_2$. These abundance measurements exist from 1965--1998. To make inference about population dynamics we use the state-space model of \citet{besbeas2002integrating},
\begin{equation}
\begin{gathered}
N_{y,t+1} = \phi_{yt} \rho_t N_{at} + \epsilon_{yt};\qquad \varepsilon_{yt} \sim \tN(0, \phi_{yt} \rho_t N_{at}),\\
N_{a,t+1} = \phi_{at} (N_{yt} + N_{at}) + \epsilon_{at};\qquad \varepsilon_{at} \sim \tN(0, \phi_{at} (1-\phi_{at}) N_{at}),\\
y_{t} \sim \tN(N_{at}, \sigma^2),
\end{gathered}
\end{equation}
where $N_{yt}$ is the number of yearling females in year $t$, $N_{at}$ is the number of adult females in year $t$, $\rho_t$ is the rate at which female offspring are produced in year $t$, and $y_t$ are the noisy observations of adult female abundance. The survival parameters are the same as the ring-recovery model, but the production is modeled on the log scale using
\[
\log (\rho_t) = \gamma_{\rho 0} + \gamma_{\rho 1} t.
\]
Adult birds are the only component of the bivariate abundance state that is (indirectly) observed, so, there is little information in these data to inform all survival and production parameters. The IPM melds the information in the two data sets to allow inference to be made on all the parameters of the joint model, 
\[
\begin{gathered} 
\bn = (\eta_{y0}, \eta_{y1}, \eta_{a0}, \eta_{a1})',  \\
\bg_1 = (\gamma_{\lambda 0}, \gamma_{\lambda 1})' \text{, and } \bg_2 = (\gamma_{\rho 0}, \gamma_{\rho 1}, \log\sigma, \log N_{y0}, \log N_{a0})'.
\end{gathered}
\]
Random unit-level $\bt$ parameters are not traditionally used, but see the analysis in Section 4.2 for an alternate version.

\subsection*{S2.2 Tabulated parameter estimates}

\begin{table}[!h]
\caption{Full results for Example II: Integrated data modelThe B\&M 2019 results were taken from Table 2 of Besbeas and Morgan (2019) and represent estimates from MCMC analysis of the full model. In the 3-stage RE model, $\bt_i \sim \tN(\bn,\sigma_{\theta}\bI)$.} \medskip
\begin{tabular}{lllll}
\hline \hline
Parameter & B\&M 2019 & 2-stage & 3-stage & 3-stage RE\\
\hline
$\eta_{y0}$ & 0.523  ( 0.067 ) & 0.519  ( 0.068 ) & 0.519  ( 0.068 ) & 0.511  ( 0.070 )\\
$\eta_{y1}$ & -0.023  ( 0.007 ) & -0.024  ( 0.007 ) & -0.024  ( 0.007 ) & -0.03  ( 0.019 )\\
$\eta_{a0}$ & 1.521  ( 0.070 ) & 1.500  ( 0.068 ) & 1.500  ( 0.068 ) & 1.496  ( 0.070 )\\
$\eta_{a1}$ & -0.028  ( 0.005 ) & -0.028  ( 0.005 ) & -0.028  ( 0.005 ) & -0.017  ( 0.014 ) \bigskip\\
$\gamma_{\lambda 0}$ & -4.563  ( 0.035 ) & -4.566  ( 0.035 ) & -4.566  ( 0.035 ) & -4.568  ( 0.035 )\\
$\gamma_{\lambda 1}$ & -0.584  ( 0.064 ) & -0.582  ( 0.065 ) & -0.582  ( 0.065 ) & -0.573  ( 0.065 )\\
$\gamma_{\rho 0}$ & -1.178  ( 0.091 ) & -1.175  ( 0.087 ) & -1.115  ( 0.086 ) & -1.128  ( 0.091 )\\
$\gamma_{\rho 1}$ & -0.425  ( 0.076 ) & -0.388  ( 0.081 ) & -0.428  ( 0.077 ) & -0.456  ( 0.087 ) \bigskip \\
$\log \sigma$ & 5.049  ( 0.136 ) & 4.884  ( 0.127 ) & 5.052  ( 0.138 ) & 4.965  ( 0.152 )\\
$\log N_{y0}$ & 5.966  ( 0.546 ) & 6.039  ( 0.447 ) & 5.985  ( 0.547 ) & 6.001  ( 0.494 )\\
$\log N_{a0}$ & 7.015  ( 0.135 ) & 7.019  ( 0.114 ) & 7.016  ( 0.137 ) & 7.019  ( 0.126 )\\
$\log \sigma_\theta$&  & & & -4.062 (0.519)\\
\hline
\end{tabular}
\end{table}

\clearpage

\bibliography{bibLibrary.bib}